\documentclass[12pt]{article}
\usepackage[margin=4cm]{caption}
\usepackage{amsmath,amsthm,amssymb,latexsym,amscd}
\usepackage{setspace}
\usepackage{subfigure, tikz, verbatim,graphics, graphicx,bm,enumerate}
\usepackage{graphicx}
\usepackage{subfigure}
\usepackage{float}
\usepackage{enumitem}
\usepackage{graphicx}
\usepackage{amsthm}
\usepackage{pgfplots}
\usepackage[english]{babel}
\usepackage{mathrsfs,esint}
\usepackage{asymptote}
\usepackage{mathtools}
\usepackage{blindtext}
\usepackage{graphicx}
\usepackage[active]{srcltx}
\usepackage{color}
\usepackage{float}
\usepackage{empheq}
 \graphicspath{ {images/} }
 \singlespace
\newtheorem*{theorem*}{Theorem}
\newtheorem*{lemma*}{Lemma}

\theoremstyle{definition}

\newcommand{\intPos}{\int_{0}^{\infty}}

\renewcommand{\epsilon}{\varepsilon}

\DeclareMathAlphabet{\mathpzc}{OT1}{pzc}{m}{it}
\pgfplotsset{compat=newest}
\definecolor{red}{rgb}{1,0,0}
\definecolor{green}{rgb}{0,1,0}
\definecolor{SeaGreen}{RGB}{46,139,87}
\definecolor{Maroon}{RGB}{128,0,0}

\title{The maximal current carried by a normal/superconducting
  interface in the absence of magnetic field} \author{L. K.
  HART\thanks{Department of Mathematics, Louisiana State University,
    Baton Rouge, LA 70803, USA.} ~ and Y. ALMOG\thanks{Department of
    Mathematics, Ort-Braude College, Carmiel 21610, Israel}}

\date{}

\begin{document}
\bibliographystyle{siam}
\maketitle
\vspace{0.5in} \textit{Abstract}: Modeling a normal/superconducting
interface, we consider a semi-infinite wire whose edge is adjacent to
a normal magnetic metal, assuming asymptotic convergence, away from the
boundary, to the purely superconducting state. We obtain that the
maximal current which can be carried by the interface diminishes in
the small normal conductivity limit.
\section{Introduction}

Consider a superconducting wire placed at a temperature lower than the
critical one.  It is well-known that at such temperatures the
superconductor loses its electrical resistivity. This means that a
current can flow through a superconducting sample and generate a
vanishingly small voltage drop. If one increases the current beyond a
certain critical threshold, the material will revert to the normal
state, even if the temperature is kept fixed below the critical level.

In the absence of magnetic field, the critical current density, at
which superconductivity is destroyed, has been obtained in the physics
literature (cf. \cite{ti96,dgpr95} and also \eqref{eq:4} below) by
neglecting the effect of boundaries. Consequently, this critical
current density does not depend at all on the normal conductivity of
the wire, a result which appears to be counter-intuitive. It is of
interest, therefore, to examine how an interface with a normal
magnetic metal affects the critical current density and to estimate
the potential drop over such an interface.

Consider, then, a superconductiong wire, denoted by $\Omega$. The wire has
interface $\partial \Omega_c$ with a magnetic metal which is at normal state.
The remaining boundary $\partial \Omega_i$ is adjacent to an insulator.  To
obtain the critical current density we use the time-dependent
Ginzburg-Landau model in the absence of a magnetic field, presented here
in the dimensionless form \cite{al12,ivko84,ruetal07} (cf. \cite{al12}
for a formal justification of the model).
\begin{subequations}
\label{eq:1}
\begin{align}
\frac{\partial \psi}{\partial t} + i \phi \psi =& \; \Delta \psi + \psi (1-|\psi|^2) & \text{in} \; \Omega \; \text{x} \; \mathbb{R}_{+} 
\\
\sigma \Delta \phi =& \; \nabla \,  \cdot [\Im(\bar{\psi}\nabla \psi)] & \text{in} \; \Omega \; \text{x} \; \mathbb{R}_{+}
\\[1.1ex]
\psi =& \;0 & \text{on} \; \partial \Omega_c \; \text{x} \; \mathbb{R}_{+}
\\
-\sigma \frac{\partial \phi}{\partial v} =& \; J & \text{on} \; \partial \Omega_c \; \text{x} \; \mathbb{R}_{+}
\\
\frac{\partial \psi}{\partial v} =& \; 0 & \text{on} \; \partial \Omega_i \; \text{x} \; \mathbb{R}_{+}
\\
\frac{\partial \phi}{\partial v} =& \; 0 & \text{on} \; \partial \Omega_i \; \text{x} \; \mathbb{R}_{+}
\\
\psi(x,0) =& \; \psi_0 &  \text{in} \; \Omega
\end{align}
\end{subequations}
In \eqref{eq:1} $\psi$ denotes the superconductivity order parameter,
which implies that $|\psi|^2$ is proportional to the number density of pairs of
superconducting electrons (Cooper pairs). Superconductors with $|\psi|=
1$ are called purely superconducting, whereas those for which $\psi=
0$ are said to be at the normal state. The scalar electric potential is
denoted by $\phi$, while the constant $\sigma$ represents the normal
conductivity of the superconducting material. In the presence of
magnetic field the normal current is given by $-\sigma(\text{A}_t + \nabla \phi)
$, where $A$ is the magnetic vector potential, but since in our case
$A = 0$ the normal current is given by $-\sigma\nabla\phi$. The
function $J: \partial \Omega_c \to \mathbb{R}$ represents the normal current
entering the sample.

The above model, with various boundary conditions, has been studied by
both physicists \cite{ivko84,ivetal82,kabe14,be15} and mathematicians
\cite{aletal15,ruetal10,ruetal10b,almog2016existence}. We mention in
particular \cite{al12} which addresses precisely the same
one-dimensional simplification we consider in the sequel.

Assuming a one-dimensional wire 
lying in $\mathbb{R}_{+}$, a stationary solution of (\ref{eq:1}) must
satisfy 
\begin{subequations}
\label{eq:2}
\begin{align}
-\psi''+i\phi \psi - \psi(1-|\psi|^2) =&\; 0  &  \text{in} \; \mathbb{R}_{+}
\\
-\sigma \phi'' + \Im[\psi'\bar{\psi}]' =& \; 0 & \text{in} \; \mathbb{R}_{+}
\\
\psi(0) =& \; 0 & -\sigma \phi'(0) = J
\\
|\psi|  \to & \rho_\infty & \text{as} \; x \to \infty
\\
\phi \to & 0  & \text{as} \; x \to \infty
\end{align}
\end{subequations}
In \eqref{eq:2}, the current $J$ is constant. The boundary conditions at
$x = 0$ represent an interface with a magnetic metal at the normal
state \cite[Eq. (4.15a)]{ti96}. As $x
\to \infty$ the sample assumes the  fully superconducting state. The latter
is given, for this simple setting (cf. \cite{al12,ti96}) by
\begin{equation}
\label{eq:3}
\psi_s = \rho_\infty e^{i\alpha x} \hspace{5mm}; \hspace{5mm} \phi \equiv 0,
\end{equation}
with $\alpha = [1-\rho^2_\infty]^{1/2}$. As the superconducting current is given
by $J=\Im[\psi'_s\bar{\psi}_s]$ we must have that
\begin{equation}
\label{eq:4}
J^2 = \rho^4_\infty(1-\rho^2_\infty).
\end{equation}
Accordingly, in \eqref{eq:2}, $J$ and $\rho_\infty$ must be related by
\eqref{eq:4}. It can be easily verified that, as \linebreak $0 \leq \rho_\infty \leq 1$, the values
of $J$ for which \eqref{eq:4} can be satisfied are limited to $J\in
[0,J_c]$ where 
\begin{displaymath}
J_c = \max_{\rho_{\infty}\in[0,1]}\rho_{\infty}^2 \sqrt{1-\rho_{\infty}} = \Big[\frac{4}{27}\Big]^\frac{1}{2}.
\end{displaymath}

Consequently, for $J=J_c$ we have
$\rho^2_\infty = 2/3$. This critical current is well known and has frequently been
documented in the literature \cite{de66,ivko84,ti96}. For $J<J_c$
(\ref{eq:4}) possesses two solutions for $\rho_\infty$. We focus interest in
this work on the solution satisfying $\rho_\infty^2>2/3$, which is conceived
in the Physics literature as the stable solution \cite{ivko84} among
the two.  

Using the polar representation $\psi = \rho e^{i \chi}$ we obtain from
(\ref{eq:2}b,c) that
\begin{displaymath}
\chi' = \frac{\sigma\phi' + J}{\rho^2},
\end{displaymath}
whenever $\rho \neq 0$. For $(\rho,\phi)$ we then obtain the following system of equations
\begin{subequations}
\label{eq:5}
\begin{align}
-\rho''+\frac{(\sigma \phi'+J)^2}{\rho^3} - \rho(1-\rho^2) &= 0 \hspace{5mm} \text{in} \hspace{1mm} \mathbb{R}_{+} &\\
-\sigma \phi'' + \rho^2 \phi &= 0 \hspace{5mm} \text{in} \hspace{1mm} \mathbb{R}_{+} \label{eq:subeq2}&\\
\rho(0) &= 0 &\\
\rho &\underset{x \to \infty} \longrightarrow  \rho_{\infty} &\\
\phi'(0) &= -\frac{J}{\sigma} &\\
\phi &\underset{x \to \infty} \longrightarrow  0
\end{align}
\end{subequations}

The present contribution focuses on the numerical evaluation of the
values of $J$ and $\sigma$ for which solutions of \eqref{eq:5} exist. As
stated above an infinite wire may admit the solution \eqref{eq:3} for
all $J\in [0, J_c]$ and positive $\sigma$. When an interface with a normal
metal at $x=0$ is added we expect that the maximal value of $J$ for
which solutions of \eqref{eq:5} exist would depend on $\sigma$.  In [1],
it is proven that the maximal value of J for which solutions of
\eqref{eq:5} can exist decays as $\sigma$ tends to zero. However as $\sigma$
gets sufficiently large, the maximal value for J asymptotically
approaches $\text{J}_c = \big[\frac{4}{27}\big]^\frac{1}{2}$

It was proven in \cite{al12} that letting
\begin{displaymath}
S(\sigma)= \{J \in \mathbb{R}_{+} \:|\:{\exists (\rho,\phi)} \in C^2(\mathbb{R}_+)\times
C^2(\mathbb{R}_{+})\hspace{1mm} \text{satisfying} \hspace{1mm}
\eqref{eq:5}\}
\end{displaymath} 
$\exists \, \text{C} \; > 0$ such that
\begin{displaymath}
\sup S(\sigma) \leq \text{C}\, \sigma^\frac{1}{4}.
\end{displaymath}
The leading order behavior as $\sigma \to 0$ has been formally obtained in
\cite{al12} as well.

The rest of this contribution is arranged as follows. In the next
section we present the numerical computation of $\sup S(\sigma)$. In \S\,3
we present the formal asymptotic expansion of $\sup S(\sigma)$ obtained in
\cite{al12} and compare it with the numerical results of \S\,2. In
addition we obtain in \S3 the potential drop over the boundary layer
(i.e. $\phi(0)$).

\section{Critical Current}

In this section we obtain the relation between the maximal current, for which
a solution of \eqref{eq:5} can exist, and $\sigma$. To this end, we need
to plot the solution $(\rho,\phi)$ of  \eqref{eq:5}.  A typical plot of
$\rho(x)$ is provided in Fig. 1 for multiple values of J and $\sigma= .2$.
\begin{figure}[H]
\begin{center}
\includegraphics[scale = .38]{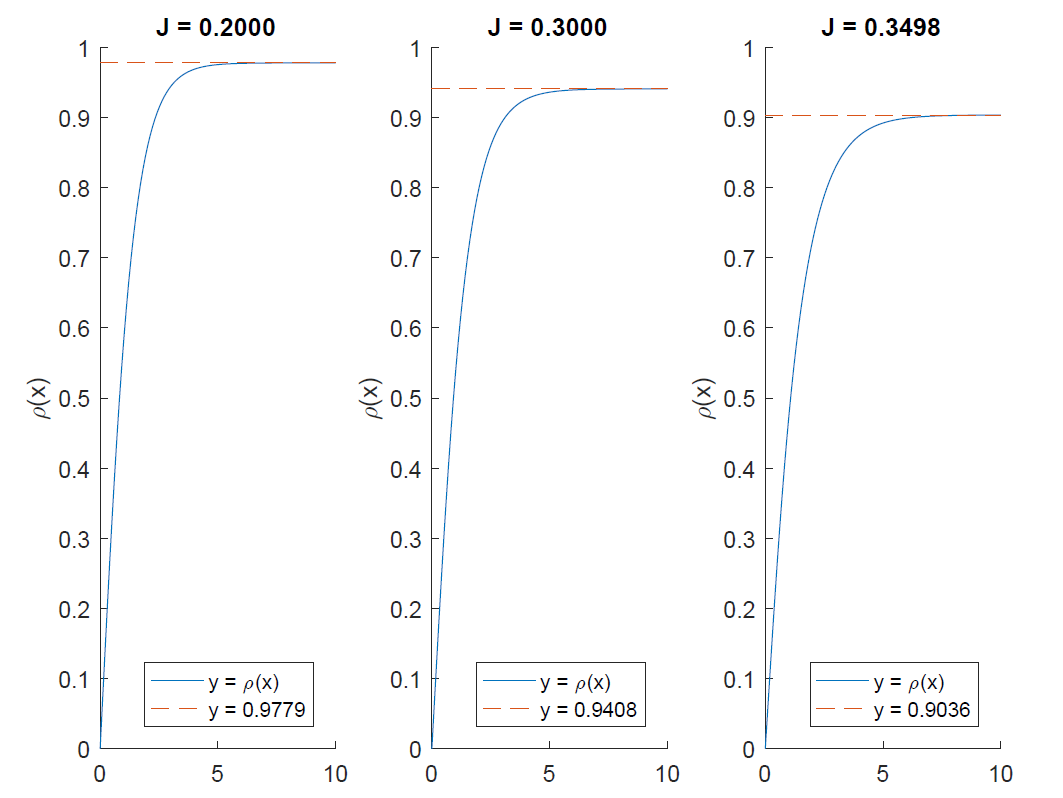}
\end{center}
\caption{Graph of $\rho$(x) in \eqref{eq:5} for  $\sigma$ =
  .2. The left two graphs asymptotically match \eqref{eq:5} as $x\to\infty$. If $J>.35$ the graph becomes nonphysical.}
\label{fig:rho}
\end{figure}
Similarly, Fig. 2 presents a plot of  $\phi$(x) for the same values, as
in Fig. 1, of $J$ and $\sigma$. \\
\\
\begin{figure}[H]
\centering
\includegraphics[scale = .37]{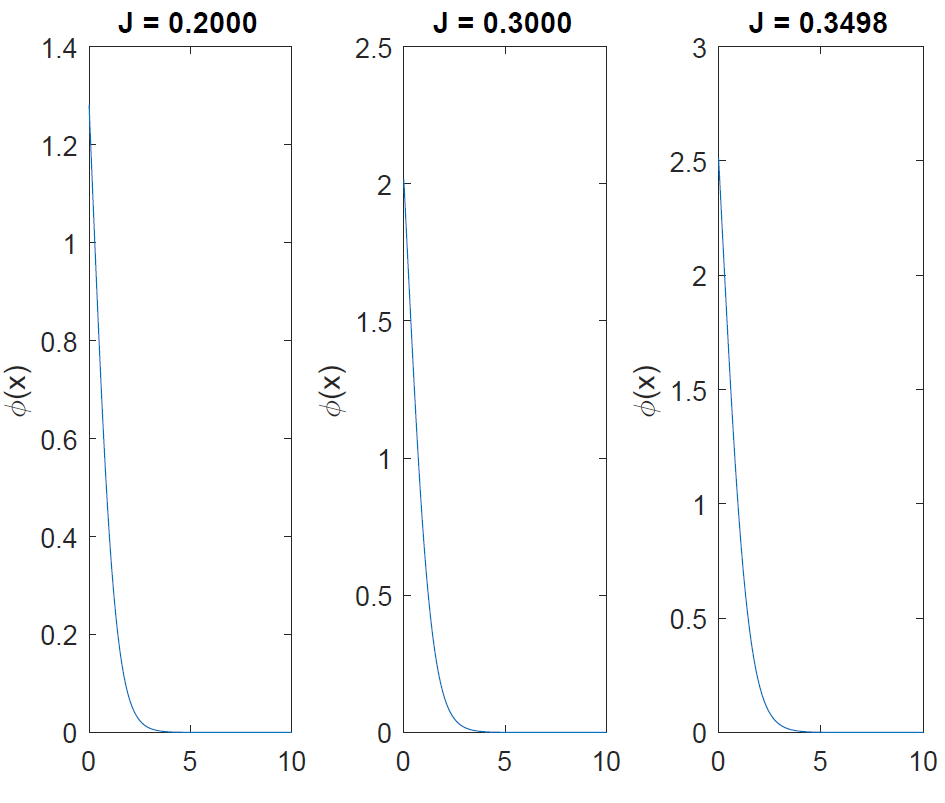}
\caption{Graph of $\phi$(x) in \eqref{eq:5} for $\sigma$ = .2. The left two graphs asymptotically match \eqref{eq:3} as $x\to\infty$. If $J>.35$ the graph becomes nonphysical.}
\label{fig:phi}
\end{figure}

We use MATLAB routine BV4PC to obtain the solution of \eqref{eq:5}. 
To this end we must first change it to a system of first order ODEs.
\begin{subequations}
\label{eq:6}
\begin{align}
f_1 =&\; \rho'
\\
f_2 =&\; \frac{(\sigma \phi + J)^2}{\rho^3} - \rho(1-\rho^2)
\\
f_3 =&\; \phi'
\\
f_4 =&\; \frac{\rho^2\phi}{\sigma}
\end{align}
with boundary conditions at $x=0$ and $x=b$ for some constant $b\gg1$.
\begin{align}
\rho(0) =&\; 0
\\
\rho(b) =&\; \rho_\infty
\\
\phi(b) =&\; 0
\\
\phi'(0) =&\; -\frac{J}{\sigma}
\end{align}
\end{subequations}

Clearly, any change in the values of $J$ and $\sigma$ will produce a
change in $(\rho,\phi)$.  To determine the maximal value of $J$, for which
a solution of \eqref{eq:5} can exist for a given $\sigma$, we increase $J$
incrementally over a set of evenly spaced numbers
$0=J_0<J_1<\ldots<J_{400}=J_c = \sqrt{\frac{4}{27}}$ (clearly,
$J_k=kJ_c/400$).  For each $J_{k}$ we graphed $\rho(x)$.  The smallest
value of $J$ for which $\rho$ does not tend asymptotically to $\rho_\infty$
(i.e. $\rho(x)$ is nonphysical), should be close to the maximal current
the wire can carry. (See Figure \ref{fig:rho} for plots of physical
solutions.) We denote this critical value by $J^{c}(\sigma)=\sup S(\sigma)$.

\section{Asymptotic Expansion}

We begin by repeating the formal asymptotic expansion, as $\sigma\to0$, of
$J^{c}(\sigma)$ from \cite{al12}. We then compare it with the numerical
solution described in the previous section.
\vspace{1ex}

Since (\ref{eq:5}b) is a Schr\"odinger equation with potential given by
$\rho^2/\sigma$, it follows, in view of (\ref{eq:5}d), that any bounded
solution decays exponentially fast as $x \to \infty$. In the limit $\sigma\to0$ we
expect the decay to take place on a fast scale. As $\rho(0)=0$, it makes
sense to assume that $\rho \sim \alpha x$
in the close vicinity of $x=0$, where $\alpha = \rho'(0).$ Note that by Lemma 2.1 in
\cite{al12}, $|\rho'| \leq \sqrt[]{\frac{2}{3}}$ and hence $\alpha$ must be
bounded as $\sigma\to0$.  The problem for
$\phi$ (\ref{eq:5}b,e,f) then takes the form
\begin{subequations}
  \begin{empheq}[left={\empheqlbrace}]{alignat=2}
    -\sigma &\phi'' + \alpha^2 x^2\phi =0 &\text{in} \; \mathbb{R}_+ \\
& \phi'(0) = -\frac{J}{\sigma} &\phi \ \xrightarrow[x \to \infty]{}0\,.
  \end{empheq}
\end{subequations}
Consider the scaled coordinate $\xi = \alpha^{\frac{1}{2}}\sigma^{-{\frac{1}{4}}}x$ and the function
\begin{displaymath}
\Phi(\xi) = \frac{\alpha^{\frac{1}{2}}\sigma^{\frac{3}{4}}}{J}\phi(x).
\end{displaymath}
The scaled form of (7) is
\begin{subequations}
\label{eq:7}
  \begin{empheq}[left={\empheqlbrace}]{alignat=2}
-&\Phi''+\xi^2 \Phi = \; 0 &\text{in} \; \mathbb{R}_+& \\
& \Phi'(0) = \; -1 & \quad \Phi \xrightarrow[\xi \to \infty]{}0\,.&
  \end{empheq}
\end{subequations}
\\

\noindent We next attempt to obtain $\alpha$ which is a priory unknown. Let 
  \begin{equation}
\label{eq:8}
    H= \frac{1}{2} \Big[ |\rho^\prime|^2 + \frac{(\sigma\phi^\prime+J)^2}{\rho^2} + \rho^2
    -\frac{1}{2}\rho^4 \Big] \,.
  \end{equation}
Differentiating (\ref{eq:8}) we obtain, with the aid of  (\ref{eq:5}a) and (\ref{eq:5}b), that
\begin{displaymath}
  H^\prime = (\sigma\phi^\prime+J)\phi \,.
\end{displaymath}
Note that the above relation is exact as it follows from (\ref{eq:5}).
Since we can precisely evaluate $H(0)$ and $H(\infty)$ in terms of $\rho_\infty$
and $\alpha$ it makes sense to approximate $H^\prime$ with the aid of
(\ref{eq:7}), and then to use the approximation to obtain an estimate of
$H(\infty)-H(0)$. Upon comparison with the exact expression we shall
be able to obtain an equation for $\alpha$.  Thus, integrating $H'$ between $0$ and
$\infty$ yields, by \eqref{eq:4}, (\ref{eq:5}c-f), and \eqref{eq:8}
\begin{displaymath}
  \int_0^\infty (\sigma\phi^\prime+J)\phi \, dx = \frac{J^2}{\alpha\sigma^{1/2}}\int_0^\infty
  (\Phi^\prime+1)\Phi\,d\xi = \rho_\infty^2 - \frac{3}{4}\rho_\infty^4-\frac{1}{2}\alpha^2 \,.
\end{displaymath}
Define 
\begin{displaymath}
  A = 2\int_0^\infty   (\Phi^\prime+1)\Phi\,d\xi \,,
\end{displaymath}
and rewrite the above equality as 
\begin{displaymath}
\frac{AJ^2}{\alpha\sigma^{1/2}} = 2\rho_\infty^2 - \frac{3}{2}\rho_\infty^4-\alpha^2.
\end{displaymath}
Since the right-hand side is bounded from above by $2$, $J$ must tend
to $0$ as $\sigma\to0$. Consequently, we must have by \eqref{eq:4} that
either $\rho_\infty\sim1$ or $\rho_\infty\sim0$. Since, as stated above, our interest
is only in the case $\rho_\infty\sim1$, we reach the asymptotic identity
\begin{displaymath}
  \alpha^2+ \frac{AJ^2}{\alpha\sigma^{1/2}} = 
\frac{1}{2} \,.
\end{displaymath}
We can now extract $J$ as a function of $\alpha$, i.e.,
\begin{displaymath}
  J^2= (\alpha-2\alpha^3)\frac{\sigma^{1/2}} {2A}\,.
\end{displaymath}
The maximum of the right-hand-side, with respect to $\alpha$, is obtained
for $\alpha=1/\sqrt{6}$. Consequently, we can conclude that the maximal
current the wire can carry is given by
\begin{equation}
  \label{eq:9}
J^c(\sigma)=\sqrt{\frac{2}{A}} \Big(\frac{1}{6}\Big)^{3/4}
  \sigma^{1/4} \,.
\end{equation}

\noindent We now attempt to obtain $A$, so we can compare
(\ref{eq:9}), obtained in  \cite{al12}, to the numerical solution of
(\ref{eq:5}).  To this end we express $\Phi$ in terms of the parabolic cylinder function $\text{U}(0,\xi)$.
It can be easily verified that
\begin{equation}
\label{eq:10}
A = -\Phi^2(0) + 2\intPos\Phi \: d\xi
\end{equation}
By \cite[Chapter 19]{abst72} we have $\Phi=C\cdot U(0,\sqrt{2}\xi)$. To obtain $C$
we utilize (\ref{eq:7}b) and write
\begin{displaymath}
\Phi'(0) = C\cdot \sqrt{2}U'(0,0) =-1 \,\implies C = \frac{-1}{\sqrt{2}U'(0,0)},
\end{displaymath}
And hence,
\begin{equation}
  \label{eq:11}
\Phi(\xi) = \frac{2^{-\frac{1}{4}}\Gamma(\frac{1}{4})}{\sqrt[]{2\pi}}U(0,\sqrt{2}\xi)\,.
\end{equation}

We estimate $\phi(x)$ using \cite[19.3.1, 19.3.3-4]{abst72} together with
\cite[19.2.5-6]{abst72} for $x\leq7$ and
\cite[19.8.1]{abst72} for $x>7$ to obtain from \eqref{eq:10}
\begin{displaymath}
A \approx 0.4336\,.
\end{displaymath}
Then, we use \eqref{eq:9} to obtain that
\begin{equation}
\label{eq:12}
J^c(\sigma)\sim 0.5602\cdot\sigma^{\frac{1}{4}}\,.
\end{equation}

For small $\sigma$, the asymptotic curve of $J^{c}(\sigma)$ aligns with the
critical J values found numerically, as can be viewed in Fig. 3 
\begin{figure}[H]
  \centering
\includegraphics[scale = .5]{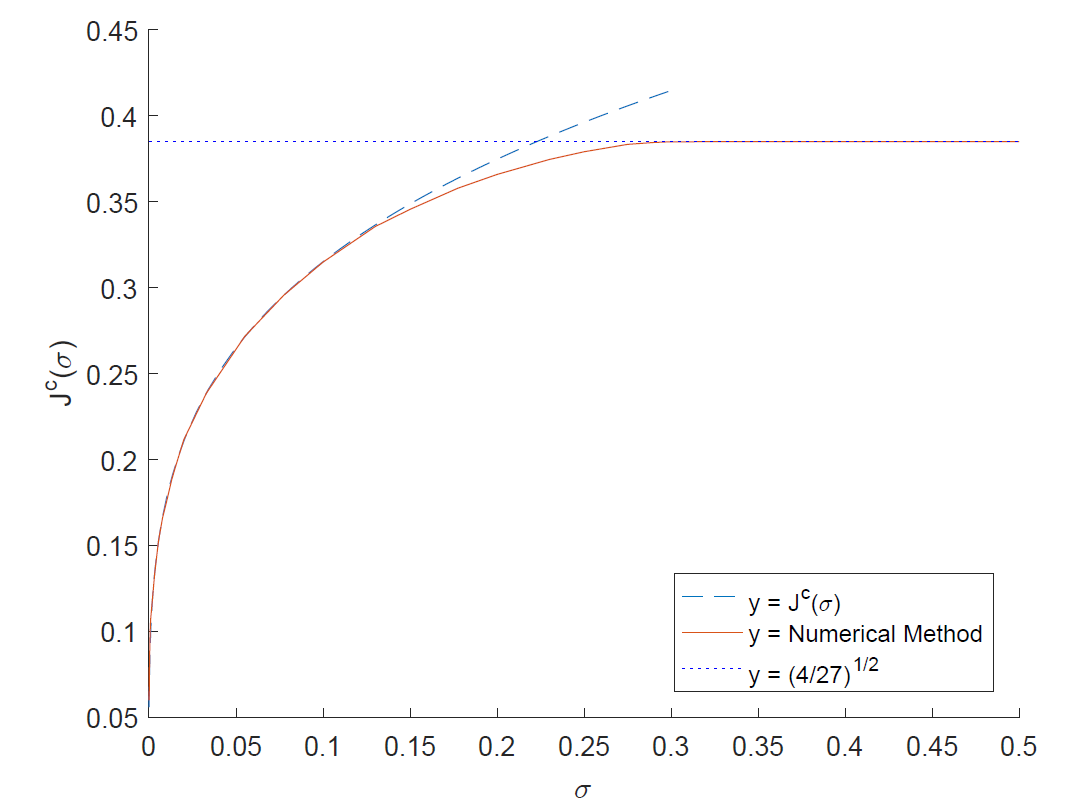}
  \caption{A plot of $J^c(\sigma)$ for $0 < \sigma \leq .5$}
  \label{fig:1}
\end{figure} 
Note that the asymptotic approximation for $J^c(\sigma)$ begins to diverge
from the numerical value at about $\sigma\approx0.13$. Such a divergence is expected
since \eqref{eq:12} cannot tend to $J_c$ as $\sigma\to\infty$.

It was established in \cite{al12} that the potential drop for $J=J^c(\sigma)$
formally satisfies, as $\sigma\to0$, 
\begin{equation}
\label{eq:13}
\phi^c(0,\sigma) \sim [3A\sigma]^{-\frac{1}{2}}\cdot\Phi(0) \approx 2.3861\cdot\sigma^{-\frac{1}{2}}\,.
\end{equation}
\\
In Fig. 4 we plot the numerical value of $\phi^c(\sigma)$ (the solid curve) and
the  asymptotic estimate given by \eqref{eq:13}.
\begin{figure}[H]
  \centering
  \includegraphics[scale=.5]{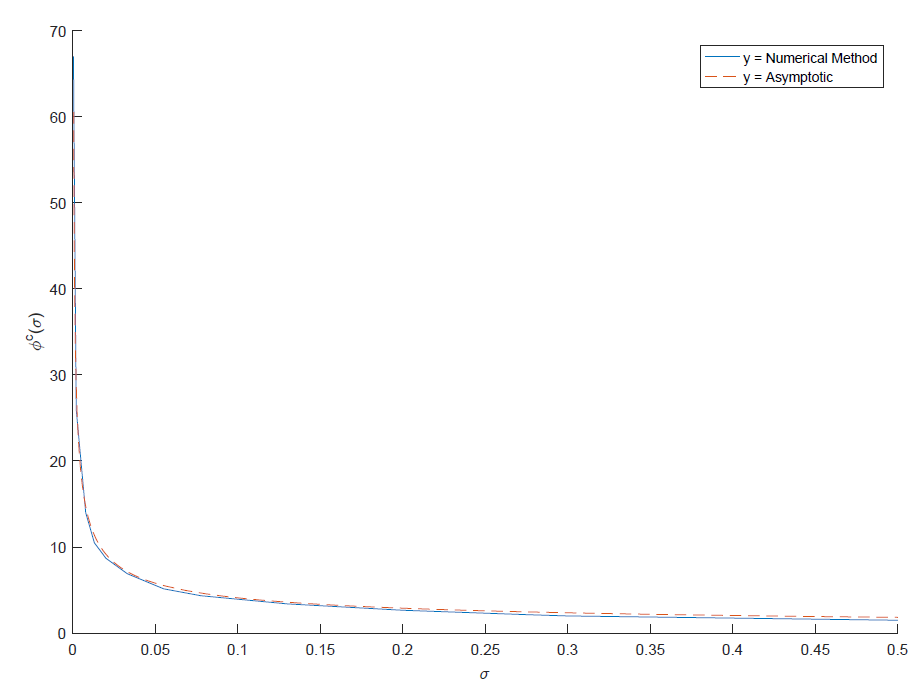}
\caption{$\phi^c(\sigma)$ for $0 < \sigma \leq .5$}
  \label{fig:2}
\end{figure}
Unlike the approximation for critical current, the asymptotic approximation for potential drop does not diverge from its numerical counterpart.
\\
\\

\section{Concluding remarks}
\label{sec:4}

In the previous sections we have obtained, for a semi-infinite
superconducting wire both the critical current at which
superconductivity is destroyed, as well as the maximal potential drop
the normal-superconducting interface can sustain. In the following we
summarize our main findings.
\begin{enumerate}
\item As $\sigma\to\infty$ the critical current tends to the asymptotic value
  $J_c=\sqrt{4/27}$ obtained in the absence of boundaries from
  \eqref{eq:4}. Accordingly, the potential drop over the interface
  tends to zero. We may conclude from here that in the large
  conductivity limit the normal-superconducting interface should not
  not have much effect on the main properties of the wire.
\item As $\sigma\to0$ the critical current $J_c(\sigma)$ diminishes like
  $\sigma^{1/4}$ whereas the potential drop diverges like $\sigma^{-1/2}$. We
  may derive from here the highly intuitive conclusion that a
  superconductor of small normal conductivity is not very useful to
  the purpose of carrying strong currents with minimal loss of energy.
\item The critical current is an increasing function of $\sigma$. Hence
  the asymptotic value $\sqrt{4/27}$ is optimal.
\end{enumerate}
{\bf Acknowledgements:}\\
This research was supported by NSF Grant DMS-1613471. 
  
\bibliography{superconductivity}
\end{document}